\documentclass[10pt,twocolumn,letterpaper]{article}

\usepackage{iccv}              % To produce the CAMERA-READY version
\definecolor{iccvblue}{rgb}{0.21,0.49,0.74}
\usepackage[pagebackref,breaklinks,colorlinks,allcolors=iccvblue]{hyperref}
\usepackage{amssymb, amsfonts}
\usepackage{mathrsfs}
\usepackage{array}
\usepackage{graphicx}
\usepackage{multirow}
\usepackage{algorithm}
\usepackage{algpseudocode}
\usepackage{xcolor}
\usepackage{amsmath}
\usepackage{amsthm}

\def\paperID{4171} % *** Enter the Paper ID here
\def\confName{ICCV}
\def\confYear{2025}

\title{Neuroverse3D: Developing In-Context Learning Universal Model for Neuroimaging in 3D}

\author{
    Jiesi Hu\textsuperscript{*1,2}, Chenfei Ye\textsuperscript{*1}, Yanwu Yang\textsuperscript{3,4}, Xutao Guo\textsuperscript{1,2}, Yang Shang\textsuperscript{1},\\
    Pengcheng Shi\textsuperscript{1}, Hanyang Peng\textsuperscript{\textdagger 2}, Ting Ma\textsuperscript{\textdagger 1, 2} \\
    {\small \textsuperscript{1}Harbin Institute of Technology, Shenzhen, China \quad
    \textsuperscript{2}Peng Cheng Laboratory, Shenzhen, China} \\
    {\small  \textsuperscript{3}University Hospital Tübingen, Tübingen, Germany \quad
    \textsuperscript{4}German Center for Mental Health (DZPG), partner site, Jena \& Tübingen, Germany} \\
    {\tt\small 22b352011@stu.hit.edu.cn, chenfei.ye@foxmail.com}\\
    {\tt\small philoso\_phy0922@163.com, tma@hit.edu.cn}
}

\begin{document}

\maketitle
% Equal contribution and corresponding author footnotes
\renewcommand{\thefootnote}{\fnsymbol{footnote}}
\footnotetext[1]{Equal contribution.}
\footnotetext[2]{Corresponding author: Hanyang Peng, Ting Ma.}
\renewcommand{\thefootnote}{\arabic{footnote}} % 恢复脚注编号样式

% 有待改进
\begin{abstract}
    In-context learning (ICL), a type of universal model, demonstrates exceptional generalization across a wide range of tasks without retraining by leveraging task-specific guidance from context, making it particularly effective for the intricate demands of neuroimaging. However, current ICL models, limited to 2D inputs and thus exhibiting suboptimal performance, struggle to extend to 3D inputs due to the high memory demands of ICL. In this regard, we introduce Neuroverse3D, an ICL model capable of performing multiple neuroimaging tasks in 3D (e.g., segmentation, denoising, inpainting). Neuroverse3D overcomes the large memory consumption associated with 3D inputs through adaptive parallel-sequential context processing and a U-shaped fusion strategy, allowing it to handle an unlimited number of context images. Additionally, we propose an optimized loss function to balance multi-task training and enhance focus on anatomical boundaries. Our study incorporates 43,674 3D multi-modal scans from 19 neuroimaging datasets and evaluates Neuroverse3D on 14 diverse tasks using held-out test sets. The results demonstrate that Neuroverse3D significantly outperforms existing ICL models and closely matches task-specific models, enabling flexible adaptation to medical center variations without retraining. The code and model weights are publicly available at \url{https://github.com/jiesihu/Neuroverse3D}.

\end{abstract}

% To develop universal models for 3D neuroimaging, two key challenges need to be addressed: (1) Existing ICL models, designed for 2D images, struggle to capture the spatial anatomical structures inherent in 3D data. (2) There is currently no memory-efficient in-context learning approach designed for 3D ICL model. 

\section{Introduction} 
\label{sec:intro} 
Computational neuroimage analysis has significantly advanced our understanding of the brain and non-invasive diagnostics, crucial for quantitative analysis and precision medicine. Recently, universal models trained on multi-domain datasets have gained attention for their ability to handle diverse tasks (e.g., segmentation, denoising, inpainting) and modalities (e.g., MRI-T1w, MRI-T2w, CT), demonstrating adaptability to domain shifts and even unseen tasks~\cite{butoi2023universeg,liu2023clip,ma2024segment,yang2024brainmass}. In-context learning (ICL) has emerged as a promising paradigm~\cite{brown2020language, wang2023images,bai2024sequential}, enabling models to adapt to domains without retraining by using image-label pairs as task-specific guidance. These informative contexts make ICL models particularly effective for segmenting tissues with intricate morphology and unifying diverse image generation tasks.

While ICL models show clinical potential in neuroimaging~\cite{butoi2023universeg,czolbe2023neuralizer,wu2024one}, they are limited by dimensionality. Existing models process 3D volumes as 2D slices, which results in the loss of inter-slice correlations and volumetric context. This lack of global spatial awareness hinders performance on neuroimaging tasks, such as hippocampus segmentation which requires 3D anatomical understanding \cite{cciccek20163d,gong20243dsam}.

To address this limitations, it is crucial to develop an ICL model that captures 3D global information. However, creating a universal ICL model with 3D input presents significant challenges, as 3D images can be over 100 times larger than their 2D slices, leading to substantial memory requirements. This results in three critical issues: (1) Context size is significantly limited, hindering performance as larger contexts size demonstrably improve outputs~\cite{butoi2023universeg, bai2024sequential, czolbe2023neuralizer, wang2023seggpt}. (2) Training models with more parameters becomes infeasible, restricting their capacity and potential performance. (3) Processing high-resolution 3D inputs becomes difficult.

\begin{figure}
\centering
\includegraphics[width=0.475\textwidth]{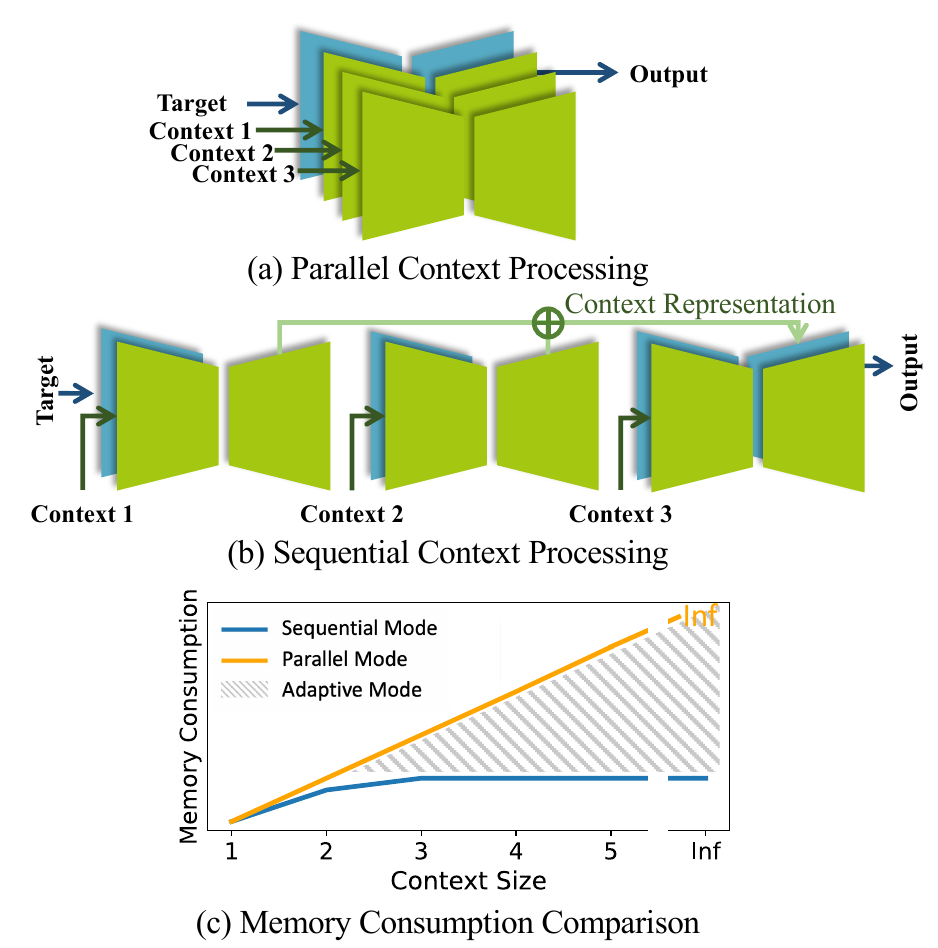}
\caption{(a) Parallel processing of all contexts, which is applied by most existing ICL models. (b) Sequential processing of each context, which greatly reduces memory usage. (c) It shows the memory increment when increasing the context size. Our proposed model processes contexts in an adaptive mode, allowing for flexible parallel and sequential processing of contexts.}
\label{fig:mainidea}
\end{figure}

To develop 3D universal models in neuroimaging, we introduce Neuroverse3D, an in-context learning model that takes 3D neuroimaging as input. To our knowledge, it is the first 3D ICL universal model for neuroimaging. To mitigate the significant memory demands, we propose Adaptive Parallel-Sequential Processing (APSP) and a U-shape fusion strategy to partition the entire context into multiple mini-contexts for processing, while ensuring the consistency of the model's gradient when training and output when inferencing. As shown in Figure \ref{fig:mainidea}, this approach can significantly reduce memory consumption, even accommodating unlimited context sizes. Additionally, to address task imbalance from increased dimensionality and enhance anatomical boundary focus, we propose an optimized loss function, further improving performance. As the first 3D ICL model in neuroimaging, we evaluate Neuroverse3D's potential across diverse tasks on held-out datasets. Our contributions are summarized as follows:

\begin{itemize}
    \item We mitigate the high memory demands of 3D image contexts using a novel APSP approach with a U-shaped fusion strategy, enabling unlimited context processing for both training and inference. This significantly improves the feasibility of deploying 3D neuroimaging ICL models in clinical practice.
    \item We develop an optimized loss function to address class imbalance resulting from increased dimensionality and enhance focus on challenging segmentation regions and anatomical boundaries, further improving performance.
    \item We propose Neuroverse3D, the first 3D ICL universal model for neuroimaging, trained on extensive multi-center datasets. Neuroverse3D significantly outperforms other ICL models, achieving over 20 absolute Dice point gains across four tasks and matching task-specific performance, offering valuable insights for universal model development.
\end{itemize}

%-------------------------------------------------------------------------
\section{Related Work}
\label{sec:related}

%% Neuroimage analysis

\subsection{Domain Challenges in Medical Imaging}
Deep learning models in medical imaging frequently face domain shifts due to heterogeneous imaging data distributions, which reduce model performance on new, unseen domains. Solutions like domain adaptation~\cite{zhang2024mapseg, yang2022source, hu2024chebyshev, wang2023fvp} require target-domain fine-tuning, limiting practical use due to the need for deep learning expertise. Domain generalization aims to produce models that generalize to new domains without fine-tuning~\cite{hu2022domain, ilse2020diva, zhang2020generalizing, ouyang2022causality}. While it alleviates the effects of domain shift, generalization remains challenging because models lack knowledge about unseen domains during training. Different from domain adaptation and domain generalization methods, universal models have emerged as a promising solution by learning generalizable image representations from a large number of datasets.

\subsection{Universal Models in Medical Imaging}
Universal models in medical imaging can be classified into three categories based on their prompt types. The first category utilizes symbolic prompts, such as points or bounding boxes, as demonstrated by models like SAM~\cite{ma2024segment, bui2024sam3d, kirillov2023segment}. The second category employs natural language prompts~\cite{liu2023clip, liu2024universal}. The third category, also known as ICL models, uses image-label pairs as prompts~\cite{czolbe2023neuralizer, butoi2023universeg, wu2024one, hu2024icl}. These universal models, trained on extensive datasets, exhibit exceptional generalization capabilities. Among these, image-label pairs provide rich contextual information, enabling ICL models to perform accurate segmentation on unseen tasks and even handle image generation tasks. However, the richness of these prompts in ICL models' inputs significantly increases memory consumption, a critical challenge that this study seeks to overcome.

\subsection{In-Context Learning}
Originally developed in natural language processing~\cite{brown2020language}, ICL vision models has recently shown potential for creating universal models that can adapt to new tasks and domains by using image-label pairs as prompts to convey task-specific information. In natural image processing, ICL-based models such as LVM~\cite{bai2024sequential}, Painter~\cite{wang2023images}, and SegGPT~\cite{wang2023seggpt} have demonstrated strong versatility across diverse tasks.

Recent studies indicate that ICL models achieve high accuracy and robust cross-domain generalization in neuroimaging, effectively addressing domain shifts across varied imaging modalities~\cite{butoi2023universeg, czolbe2023neuralizer, wu2024one, hu2024icl}. Models like UniverSeg~\cite{butoi2023universeg}, Neuralizer~\cite{czolbe2023neuralizer}, and One-Prompt~\cite{wu2024one} leverage context sets to adapt to new domains and tasks without retraining, performing effectively in few-shot scenarios. However, current ICL models are unable to directly process 3D neuroimage data, limiting their ability to fully capture the volumetric information of 3D images.

\section{Method}
\label{sec:method}

\begin{figure}
\centering
\includegraphics[width=0.475\textwidth]{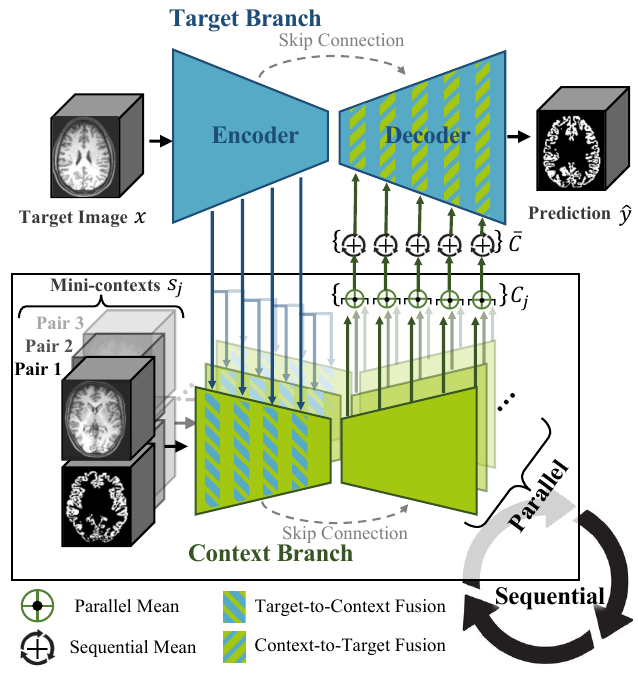}
\caption{Illustration of our model architecture. The network consists of two branches for extracting representations from the target and context images, respectively. The Target-to-Context and Context-to-Target Fusion modules enable information exchange between the branches. Our model adaptively performs both parallel and sequential context processing.}
\label{fig:framework}
\end{figure}

In-Context Learning (ICL) models for neuroimaging achieve universality by learning task-specific processing from context examples, typically image-label pairs. In the following sections, we detail our Adaptive Parallel-Sequential Processing (APSP) and U-shaped fusion strategy for efficient context processing, followed by our optimized loss function for further performance improvements.

\begin{figure}
\centering
\includegraphics[width=0.35\textwidth]{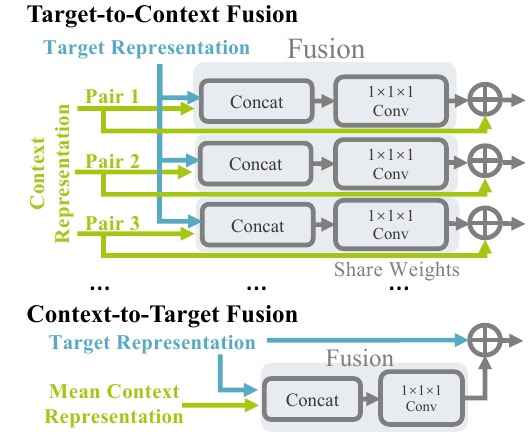}
\caption{Illustration of the Target-to-Context Fusion and Context-to-Target Fusion modules.}
\label{fig:fusion}
\end{figure}

\subsection{Model} 
As illustrated in Figure \ref{fig:framework}, the model contains two 3D U-Net branches: a target branch for extracting target image representations and a context branch for extracting context representations. These branches communicate through the U-shape fusion strategy at each stage. The context branch receives concatenated image-label pairs as input and uses APSP to compute the mean context representation, which is then passed to the target branch decoder for final prediction generation.

\vspace{0.4em}\noindent\textbf{Context Partition.} 
Given a task-specific context $S = \{(x_i, y_i)\}_{i=1}^L$, where each $(x_i, y_i)$ represents an image-label pair within a context set of size $L$, we perform context partitioning. This divides $S$ into $n = \lceil \frac{L}{\ell} \rceil$ disjoint mini-contexts $\{s_j\}_{j=1}^n$, each containing $\ell$ image-label pairs, with $\ell$ determined by available memory resources. Larger values of $\ell$ decrease inference time but increase memory usage, and vice versa. If $L$ is not divisible by $\ell$, the final mini-context will contain fewer than $\ell$ pairs, ensuring that $\bigcup_{j=1}^n s_j = S$.

\vspace{0.4em}\noindent\textbf{Adaptive Parallel-Sequential Processing.}
APSP accommodates arbitrary context partitions. Specifically, each mini-context is initially processed in parallel, and the resulting context representations are subsequently fused during sequential processing.

During parallel processing, all image-label pairs within a mini-context $s_j$ are processed simultaneously through shared-weight 3D U-Nets in the context branch. The context representation for each mini-context $s_j$ is computed as $C_j = g(s_j, f_{\text{enc}}(x))$, where $x$ represents the target image, $g$ denotes the context branch, and $f_{\text{enc}}$ is the encoder of the target branch. Here, $C_j$ is a set comprising the representations from each stage of the context decoder, defined as $C_j = \{\bar{c}_{\text{dec}, j}^i\}_{i=1}^{\mathcal{D}}$, where $\mathcal{D}$ represents the number of decoder stages. $\bar{c}_{\text{dec},j}^i$ is computed via a parallel mean aggregation of image-label pairs inside $s_j$:

\begin{equation}
    \bar{c}_{\text{dec},j}^i = \frac{1}{\ell} \sum_{k=1}^{\ell} c_{\text{dec}, j, k}^i,
\end{equation}
where $c_{\text{dec}, j, k}^i$ denotes the output representation of the $k$-th image-label pair from the $i$-th decoder stage in mini-context $s_j$. This mean aggregation ensures the output is invariant to the order of image-label pairs within a mini-context.

\begin{algorithm}
\caption{Adaptive Parallel-Sequential Processing. Let $x$ and $S$ denote the target image and context set, respectively. $\bar{c}_{\text{dec}, j}^i$ represents the context representation of the $j$-th mini-context at the $i$-th decoder stage. $\text{len}(s_j)$ indicates the number of image-label pairs in $s_j$.}\label{alg:mini_context_update}
\begin{algorithmic}
    \State $(x, y, S) \sim \mathcal{T}$ \Comment{Sample data}
    \State $\{s_j\}_{j=1}^n \gets S$ \Comment{Split context set into mini-contexts}
    \State $\{\bar{c}_{\text{dec}}^i\}_{i=1}^{\mathcal{D}} \gets \{0\}_{i=1}^{\mathcal{D}}$ \Comment{Initialize representation}
    \State $ w \gets 0$ \Comment{Initialize weight accumulator}
    \For{$j = 1, \ldots, n$} \Comment{Sequential processing}
        \State $\{\bar{c}_{\text{dec}, j}^i\}_{i=1}^{\mathcal{D}} = g(s_j, f_{\text{enc}}(x))$ %\Comment{Compute representations}
        \State $\alpha \gets \frac{w}{w + \text{len}(s_j)} $
        \State $\{\bar{c}_{\text{dec}}^i\}_{i=1}^{\mathcal{D}} \gets \{\alpha\bar{c}_{\text{dec}}^i+(1-\alpha)\bar{c}_{\text{dec},j}^i\}_{i=1}^{\mathcal{D}}$ \Comment{Sequential mean}
        \State $w \gets w + \text{len}(s_j)$ \Comment{Update cumulative weight}
        \State \textbf{delete} $\{\bar{c}_{\text{dec},j}^i\}_{i=1}^{\mathcal{D}}$ \textbf{if} $j \neq n$ \Comment{Release memory}
    \EndFor
    \State $\bar{C} \gets \{ \bar{c}_{\text{dec}}^i \}_{i=1}^{\mathcal{D}}$
    \State $\hat{y} \gets f_{dec}(f_{\text{enc}}(x), \bar{C})$ \Comment{Generate final prediction}
\end{algorithmic}
\end{algorithm}

The sequential computation procedure for $\bar{C}$ is outlined in Algorithm~\ref{alg:mini_context_update}. The model processes each mini-context sequentially, updating $\bar{C}$ iteratively. The sequential mean operation ensures equal weighting of image-label pairs across the entire context set, making $\bar{C}$ invariant to the ordering of mini-contexts. After $\bar{C}$ is obtained, and the final prediction is computed as $\hat{y} = f_{\text{dec}}(f_{\text{enc}}(x), \bar{C})$, where $f_{\text{dec}}$ denotes the decoder of the target branch. The combination of parallel and sequential mean operations in APSP guarantees consistent mean context representation computation for any context order or partition.

To save memory, intermediate features from processed mini-contexts are discarded after contributing to $\bar{C}$, with only the last mini-context retained for gradient computation during training. This reduces memory usage to the final mini-context rather than the entire context set. Moreover, during training, we randomly shuffle mini-contexts and scale the last mini-context's representation $C_n$ by a factor of $n$ during gradient computation. This ensures the expected value of the gradients computed using APSP is equivalent to the expectation when no mini-contexts are discarded, as demonstrated below:
\begin{equation}
    \mathbb{E}_\pi \left[ \nabla \mathcal{L}_{\text{APSP-scaled},\pi} \right] = \nabla \mathcal{L}_{\text{full}},
\label{eq:gradient_equ}
\end{equation}
where $\nabla \mathcal{L}_{\text{full}}$ and $\nabla \mathcal{L}_{\text{APSP-scaled},\pi}$ denote the gradients obtained from full-context processing and APSP, respectively. $\pi$ represents the original index of the mini-context selected for gradient computation in APSP. A proof of this equivalence is provided in the supplemental \cref{sec:Gradient_Equivalence}.

\vspace{0.4em}\noindent\textbf{U-Shape Fusion Strategy.} The U-shaped design makes feature propagation through two phases: first, from the target branch to the context branch at each encoder stage via Target-to-Context Fusion modules, and then from the context branch back to the target branch at each decoder stage through Context-to-Target Fusion modules, as illustrated in \cref{fig:framework}. This U-shaped representation flow is actually essential, as it allows the model to avoid storing features from previously processed mini-contexts, a limitation of alternative fusion strategies. Further explanation are provided in the supplemental \cref{sec:U_Shape}.

As illustrated in \cref{fig:fusion}, the fusion components within the Target-to-Context Fusion and Context-to-Target Fusion modules are identical. The fusion operation is defined as 
\begin{equation}
    \text{Fusion}(c^i, t^i) = \text{Conv}(c^i \,||\, t^i),
\end{equation}
where $c^i$ and $t^i$ denote the feature representations of the $i$-th stage from the context and target branches, respectively. $\text{Conv}$ represents a convolutional operation, and $||$ signifies feature map concatenation.

\vspace{0.4em}\noindent\textbf{Target-to-Context Fusion.} This module is incorporated after each encoder stage of the context branch as follows:
\begin{equation}
\hat{c}^i = \mathcal{A}(\text{Fusion}(c^i, t^i) + c^i),
\end{equation}
where $\hat{c}^i$ represents the fused context representation propagated to the next stage, and $\mathcal{A}$ denotes the activation function. Convolutional parameters are shared across all image-label pairs to enable parallel processing.

\vspace{0.4em}\noindent\textbf{Context-to-Target Fusion.} This module is integrated after each decoder stage of the target branch as follows:
\begin{equation}
\hat{t}^i = \mathcal{A}(\text{Fusion}(\bar{c}^i, t^i) + t^i),
\end{equation}
where $\hat{t}^i$ denotes the fused target representation that is sent to the next decoder stage, and $\bar{c}^i$ represents the mean context representation.

\subsection{Loss Function}
The total loss function is defined as: 
\[
\mathcal{L}_{\text{total}} = \mathbb{E}_{\tau} \left[ \mathbb{E}_{(x^{\tau}, y^{\tau}), S^{\tau}} \left[ \lambda^{\tau} \mathcal{L}^{\tau}(\hat{y}^{\tau}, y^{\tau}) \right] \right],
\]
where $\tau$ denotes the sampled tasks, and $\lambda^\tau \in \mathbb{R}^+$ represents the task-specific weighting coefficient. The procedure begins by sampling a task $\tau$ from all tasks, followed by sampling a target image $x^\tau$, corresponding ground truth $y^{\tau}$ and a support set $S^\tau$ within that task. The task-specific loss function $\mathcal{L}^\tau$ is then applied to compute the loss between the $\hat{y}^\tau$ and $y^\tau$.

For both segmentation and generation tasks, we employ an $L_1$-based loss function, with modifications to accommodate the characteristics of 3D neuroimages. 

Specifically, 3D neuroimage segmentation, unlike its 2D counterpart, exhibits increased sparsity in structures like the hippocampus and amygdala due to the added dimension, resulting in greater class imbalance. To mitigate this, we propose a modified $\mathcal{L}_{\text{smooth}-L_{1}}$ loss:

\begin{equation}
    \mathcal{L}^{\text{seg}}(\hat{y}, y) = 
    \begin{cases} 
    \frac{1}{3}|\hat{y} - y|^3, & \text{if}\ |\hat{y} - y| < 1, \\
    |\hat{y} - y| - \frac{2}{3}, & \text{otherwise}.
    \end{cases}
\end{equation}

This formulation, using a higher exponent than standard $\mathcal{L}_{\text{smooth}-L_{1}}$, prioritizes challenging anatomical regions over simpler ones, balancing performance across diverse tasks. We opted against Dice or Focal loss to avoid further complicating the balance between segmentation and generation tasks.

For generation tasks, we apply a standard $\mathcal{L}_{\text{smooth}-L_{1}}$ loss. Furthermore, to achieve better generative performance on brain images filled with intricate anatomical boundaries, we incorporate an additional $\mathcal{L}_{\text{smooth}-L_{1}}$ loss on the intensity difference of the image, inspired by~\cite{liu2023brain}:
\begin{equation}
    \mathcal{L}^{\text{gen}}(\hat{y}, y) = \frac{\mathcal{L}_{\text{smooth}-L_{1}}(\hat{y}, y)}{2} + \frac{\mathcal{L}_{\text{smooth}-L_{1}}(\Delta \hat{y}, \Delta y)}{2},
\end{equation}
where $\Delta y$ denotes the intensity difference at the current voxel relative to its neighboring voxels.

% During training, gradients for the target and context branches are computed only for the final mini-context. As outlined in Algorithm \ref{alg:mini_context_update_}, features and gradients from all but the final mini-context are discarded. This reduces memory usage by avoiding gradient computation for each mini-context. While this approach supports an unlimited context size $L$ with a fixed mini-context size $\ell$, the sequential processing increases training time. However, it enables the model to adapt to different context sizes $L$ during inference.

% During training, gradients for the target branch and the context branch are computed only for the final mini-context. As shown in Algorithm \ref{alg:mini_context_update_}, features and their corresponding gradients from all mini-contexts, except for the final one, are deleted, retaining only the updated $\bar{c}^i$. This approach reduces memory usage by avoiding gradient computation for every mini-context. Thus, provided the mini-context size $\ell$ remains constant, this method allows for an unlimited context size $L$, though at the cost of increased training time because of the sequential context processing. This enables the model to train with various context sizes $L$, ensuring adaptability to different $L$ values during inference.

%-------------------------------------------------------------------------

\section{Experiments}
\subsection{Data}
\noindent\textbf{Datasets.} To ensure robust cross-center generalization and diversity, we collected 19 datasets with multiple modalities and centers, comprising 43,674 scans. The dataset includes common neuroimaging modalities including T1, T2, FLAIR, MRA, DWI, ADC, PD, and CT. Data from 15 datasets, including 38,126 scans and 5,632 segmentation masks, were used for training and validation~\cite{yang2023benchmarking, CAS2023, hernandez2022isles, liew2022large, IXI, hoopes2022learning, marcus2007open, flanders2020construction, ADHD, jack2008alzheimer, alexander2017open, holmes2015brain, gera2023characterizing, nugent2022nimh, sudlow2015uk, menze2014multimodal}, with a random 9:1 split. Data from the remaining 4 datasets, including 5,548 images and 1,096 segmentation masks, was used as held-out datasets~\cite{kuijf2019standardized, liu2021chinese, FCON, marek2011parkinson} to assess performance on unseen centers. The held-out dataset was divided into an 8:2 split for the meta context set (from which ICL models select context) and the test set. During training, we used both the original images and images aligned to the MNI 152 template space~\cite{fonov2009unbiased} to increase data diversity. We utilized FreeSurfer~\cite{fischl2012freesurfer} to generate additional anatomical segmentation labels for 3 datasets~\cite{gera2023characterizing, nugent2022nimh, FCON}.

\vspace{0.4em}\noindent\textbf{Data Preprocessing.} To mitigate variability in image size across neuroimaging datasets, we first resampled the voxel resolution to $1\,\text{mm}^3$. The brain images were then rescaled to fit within a $128 \times 128 \times 128$ 3D image. Intensity values were normalized to the [0, 1] range based on the 0.02 and 0.98 intensity percentiles. Segmentation masks were binarized by assigning 0 to the background and 1 to the foreground.

\vspace{0.4em}\noindent\textbf{Neuroimaging Tasks.} With these datasets, our model was trained on multiple segmentation tasks, including anatomical segmentation~\cite{billot2020learning}, tumor segmentation~\cite{menze2014multimodal}, vessel segmentation~\cite{yang2023benchmarking}, and generation tasks, including bias field correction~\cite{goldfryd2021deep}, inpainting~\cite{liu2021symmetric}, super-resolution~\cite{laguna2022super}, Gaussian noise removal, salt-and-pepper noise removal~\cite{laguna2022super}, 2D-to-3D reconstruction~\cite{iglesias2021joint}, modality transformation~\cite{osman2022deep}, and skull stripping~\cite{hoopes2022synthstrip}. For segmentation tasks, binary masks were produced by thresholding the model outputs at 0.5.

\vspace{0.4em}\noindent\textbf{Image and Task Augmentation.} During training, we applied data augmentation across two dimensions: image and task. In addition to standard image augmentations, we enhanced the contrast diversity of images using a randomly initialized convolutional network~\cite{ouyang2022causality}. For task augmentation, we introduced random task overlapping and also followed the method introduced in~\cite{czolbe2023neuralizer}. Synthetic brain data with random modalities~\cite{billot2023synthseg} was also added. Further details are provided in supplemental \cref{sec:aug}.

Detailed information on datasets and tasks is provided in the supplemental \cref{sec:data_sup}.

\begin{figure*}
\centering
\includegraphics[width=0.99\textwidth]{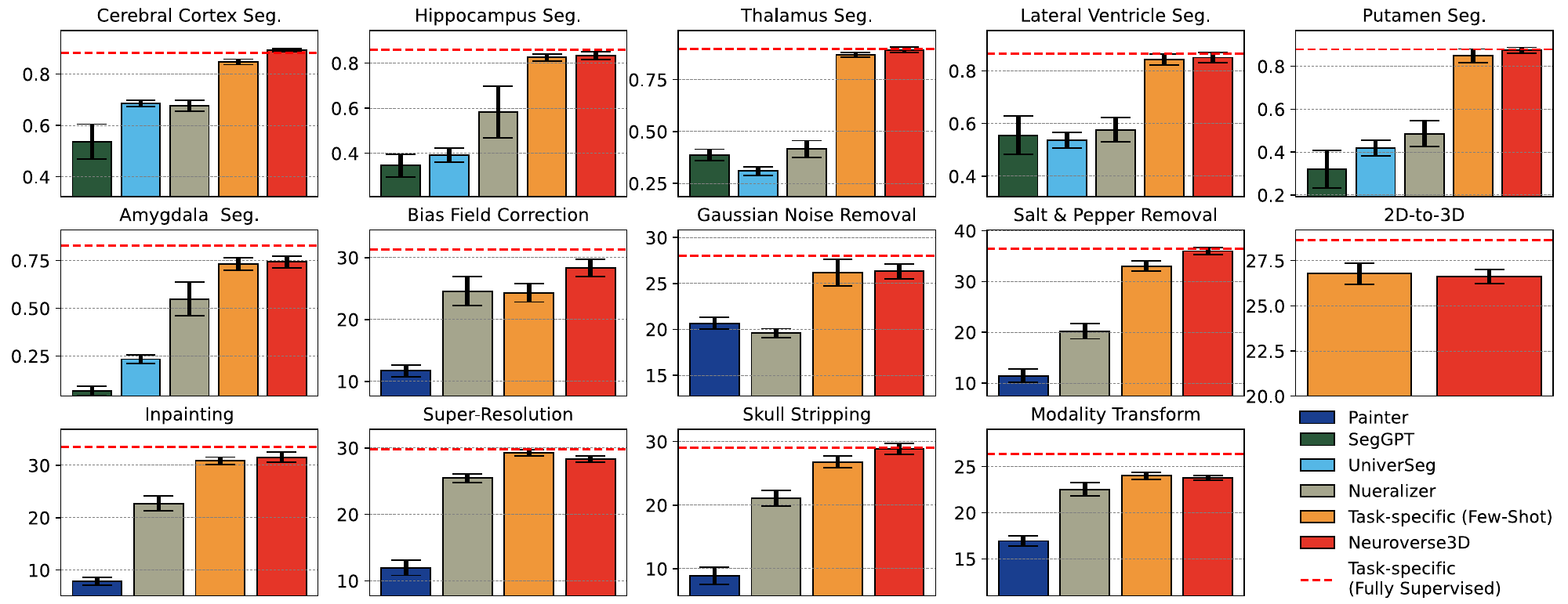}
\caption{Performance comparison of Neuroverse3D with other models on held-out datasets, under a context size of 8. This includes ICL models trained on neuroimages (UniverSeg~\cite{butoi2023universeg}, Neuralizer~\cite{czolbe2023neuralizer}), models trained on natural images (Painter~\cite{wang2023images}, SegGPT~\cite{wang2023seggpt}), and task-specific models in few-shot and fully supervised settings. The Dice coefficient is used for segmentation tasks, and PSNR for generation tasks.}
\label{main_fig}
\end{figure*}

\subsection{Compared Models}
\noindent\textbf{Neuroverse3D.} The target and context branches are based on a 5-stage 3D U-Net architecture~\cite{cciccek20163d}, with each stage comprising two residual blocks~\cite{he2016deep} constructed from $3 \times 3$ convolutional layers. The network initiates with 32 channels in the first stage, subsequently doubling the channel count at deeper stage. GELU activation functions~\cite{hendrycks2016gaussian} are employed throughout the network. Neuroverse3D is trained jointly across all tasks, whereas Neuroverse3D-unseen maintains an identical architecture but excludes the evaluated task during training. Consequently, we trained a group of Neuroverse3D-unseen models for unseen task evaluation.

\vspace{0.4em}\noindent\textbf{Task-Specific Models.} We compared Neuroverse3D against 3D task-specific models, which shared identical architecture and channel configurations but lacked the context branch. These task-specific models were trained on the held-out dataset, thereby mitigating domain shift concerns. This approach simulates the scenario where medical centers traditionally train models for specific scenario without employing ICL model. Identical data augmentation strategies were applied, excluding task augmentations that could potentially disadvantage task-specific models, such as the application of Sobel filters to segmentation masks. We evaluated performance under both few-shot learning scenarios, where models were trained with only context set data, and fully supervised learning scenarios, leveraging all available data from the meta-context set.

\vspace{0.4em}\noindent\textbf{Other In-Context Learning Models.} We compared our method with state-of-the-art ICL methods, including Painter~\cite{wang2023images}, SegGPT~\cite{wang2023seggpt}, UniverSeg~\cite{butoi2023universeg}, and Neuralizer~\cite{czolbe2023neuralizer}, all designed for 2D inputs. To accommodate the 3D context, we randomly sampled axial slices containing the segmentation target or brain to construct the 2D context set. The 2D context sizes were set to 1 for Painter~\cite{wang2023images}, 8 for SegGPT~\cite{wang2023seggpt}, 32 for Neuralizer~\cite{czolbe2023neuralizer}, and 64 for UniverSeg~\cite{butoi2023universeg}, corresponding to the reported optimal context sizes in their respective publications. The target 3D images were similarly decomposed into axial 2D slices and fed into these methods. The resulting 2D outputs were then reassembled into 3D images for metric evaluation. We downloaded the corresponding pretrained weights for these models to perform inference. UniverSeg~\cite{butoi2023universeg} and SegGPT~\cite{wang2023seggpt} are restricted to segmentation tasks, and all 2D models are incapable of performing 2D-to-3D reconstruction task.

Comprehensive details regarding the training and evaluation protocols are provided in supplemental \cref{sec:exp_sup}.

\begin{figure*}
\centering
\includegraphics[width=0.99\textwidth]{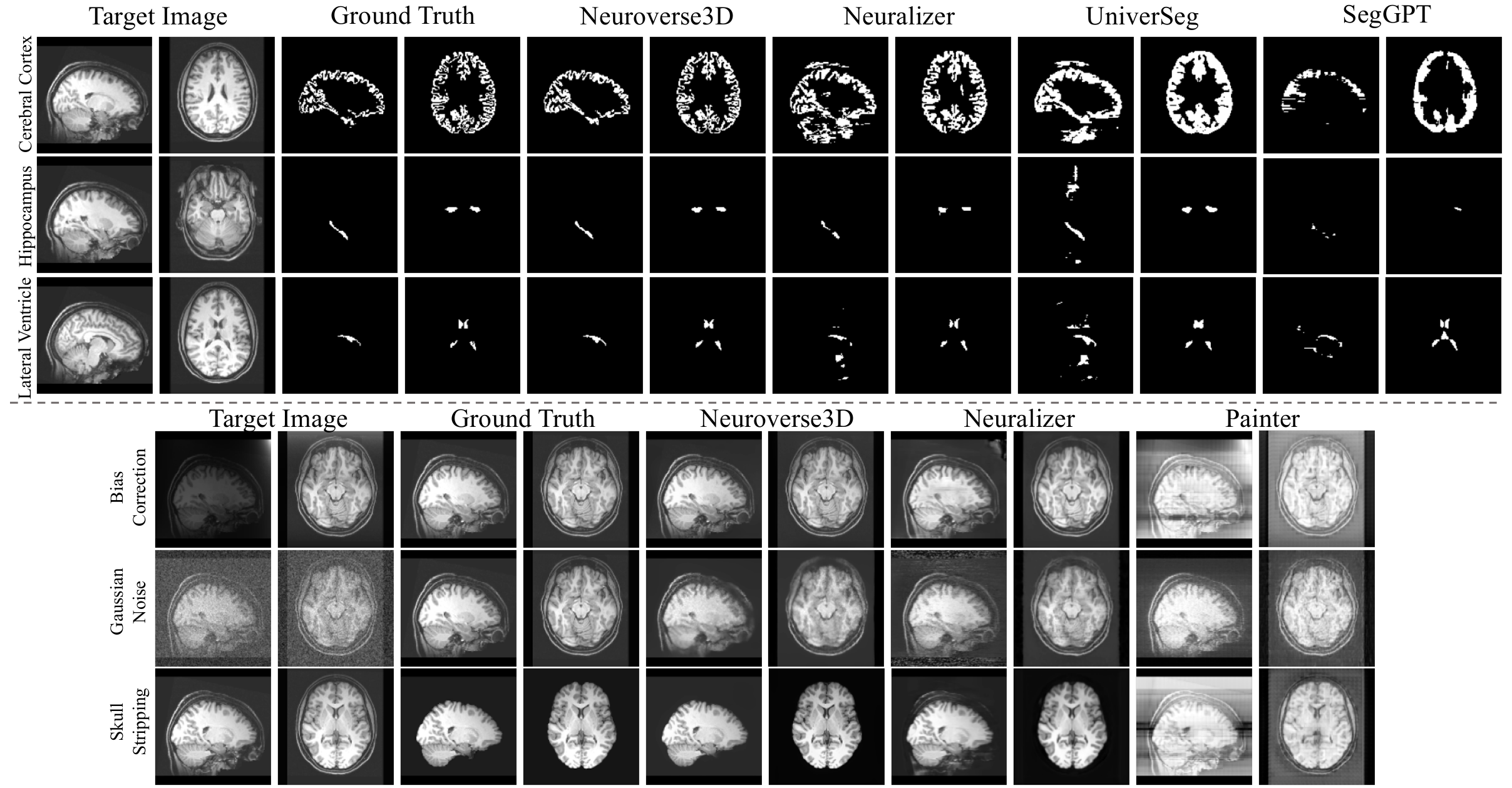}
\caption{Qualitative results for four representative tasks are presented. To comprehensively illustrate the 3D image results, slices from both sagittal (first row) and axial (second row) views are displayed. Qualitative results for all tasks can be found in supplementary \cref{sec:Qualitative_Results}}
\label{fig:quantitative}
\end{figure*}

\subsection{Model Comparison on Held-Out Datasets}
As demonstrated in \cref{main_fig}, we evaluated Neuroverse3D against SOTA ICL models and task-specific models. Comparisons with Painter were excluded for segmentation tasks due to its low Dice scores.

In segmentation tasks, Neuroverse3D significantly outperformed all other ICL models, with Dice score gains exceeding 20 percentage points for targets such as hippocampus, thalamus, lateral ventricle, and putamen. Furthermore, it surpassed the performance of few-shot models in all segmentation tasks, demonstrating that employing an ICL model in few-shot scenarios offers medical centers a more cost-effective and accurate solution. Moreover, the performance of our model closely approached that of fully supervised U-Net models. We attribute Neuroverse3D's substantial performance improvement primarily to its effective utilization of a 3D architecture, enabling it to not only better process 3D target images but also to capture the global information from the 3D context set, facilitating a more precise understanding of the segmentation targets.

In generation tasks, Neuroverse3D outperformed all other ICL models, surpassing Neuralizer and Painter across all tasks. Its performance closely matched few-shot models and nearly reached fully supervised levels in salt-and-pepper noise removal and skull stripping. This highlights the potential of 3D ICL models in generation tasks. Notably, our model's performance gain in some generation tasks was less significant. This may be attributed to certain tasks, like modality transform and super-resolution, where 2D context sufficiently conveys task information, reducing the advantage of a 3D architecture.

\cref{fig:quantitative} illustrates qualitative predictions of ICL models, showcasing results for both axial and sagittal slices. As 2D ICL models generate predictions on axial slices, discontinuities are frequently observed on sagittal slices. In segmentation, compared to Neuroverse3D, other models demonstrate lower accuracy and frequent false positives/negatives in sagittal slices. In generation tasks, Neuralizer generates excessive background errors, while Painter, despite reasonable performance on Gaussian denoising, struggles with neuroimage-specific tasks like bias correction and skull stripping. Conversely, Neuroverse3D maintains consistent accuracy across all tasks.

These results demonstrate Neuroverse3D's accurate performance across diverse segmentation and generation tasks from different medical centers, attributed to its novel approach for 3D context processing and optimized loss function.

\begin{figure}
\centering
\includegraphics[width=0.47\textwidth]{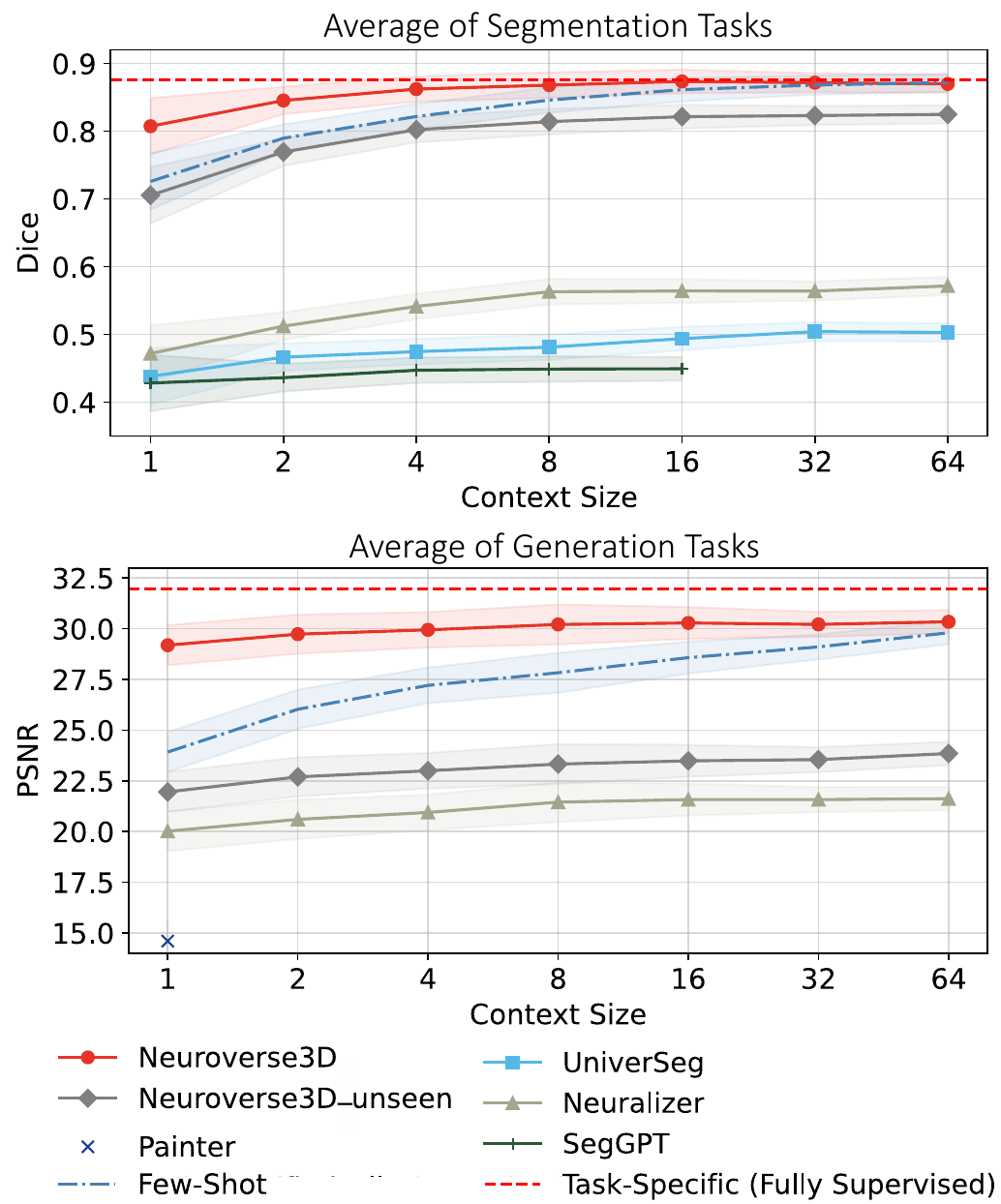}
\caption{Performance across different 3D context sizes. A group of Neuroverse3D-unseen models was trained to assess performance on tasks not encountered during training. Due to computational constraints, segmentation task evaluations here only include cerebral cortex, hippocampus, thalamus, and lateral ventricle. Generation task evaluations include bias field correction, Gaussian noise removal, and salt-and-pepper noise removal.}
\label{context_size}
\end{figure}

\subsection{Impact of Context Size}
Figure \ref{context_size} illustrates model performance across varying context sizes. Our model consistently outperforms other models across all context sizes in both segmentation and generation tasks. Besides, Neuroverse3D, along with other models, shows a significant performance increase with larger context sizes, emphasizing the critical role of context quantity for ICL models. The average metric improvement resulting from increased context size diminishes as the context size increases, reaching a plateau around 16. However, the standard deviation of predictions continues to decrease, leading to more stable predictions. APSP overcomes memory limitations, enabling Neuroverse3D to support larger or even unlimited context sizes under limited resources, thereby facilitating the practical use of Neuroverse3D. SegGPT, due to memory limitations, could only support a maximum context size of 16 in our experiments.

\subsection{Performance on Unseen Tasks}
In \cref{context_size}, Neuroverse3D-unseen shows the model’s ability to generalize to tasks not encountered during training, which is of clinical interest.

For segmentation tasks, Neuroverse3D-unseen exhibited a slight performance reduction compared to Neuroverse3D, with approximately a 5-point decrease in the Dice coefficient after the context size was greater than or equal to 8. However, it still significantly outperformed other ICL models. For generation tasks, Neuroverse3D-unseen showed a more substantial performance decline, although it remained marginally superior to Neuralizer. We attribute the greater performance drop in generation tasks compared to segmentation tasks to the fact that, during training, Neuroverse3D-unseen encountered a wider variety of segmentation tasks, including segmentations of over thirty brain structures and random combinations of brain structures. This led to a learning process for segmentation tasks closer to meta-learning, enabling cross-task generalization. In contrast, the training set contained fewer generation tasks, leading the model to tend to memorize each generalization task. This suggests that incorporating a greater diversity of tasks into ICL models is a promising direction for further enhancing cross-task generalization capabilities.

\subsection{Memory and Time Requirements}
\begin{figure}
\centering
\includegraphics[width=0.4\textwidth]{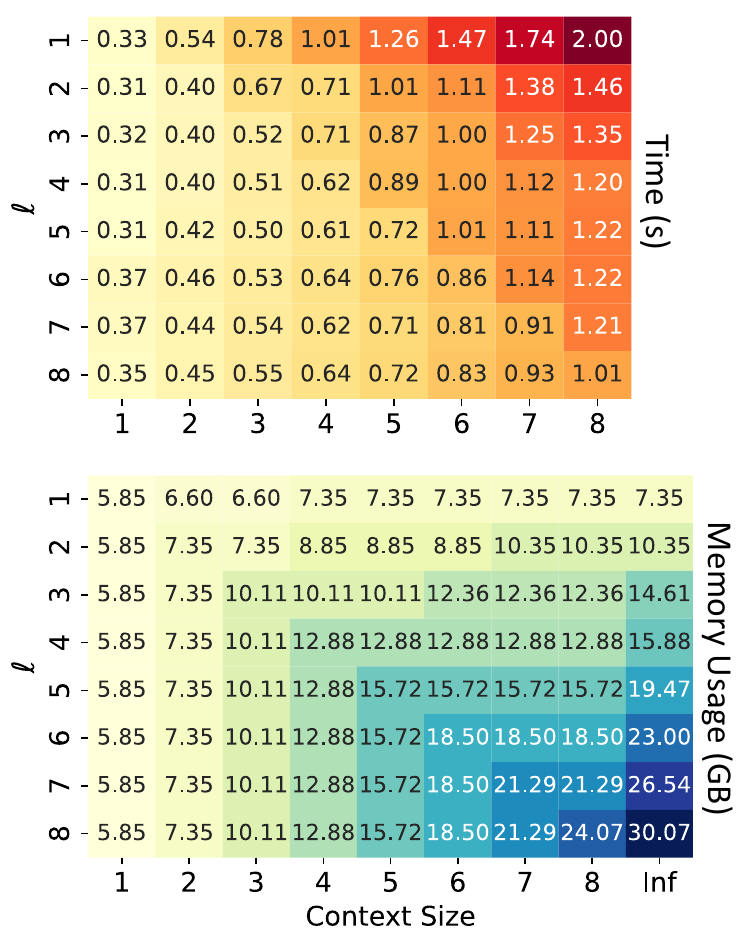}
\caption{Time and memory consumption for different context sizes and mini-context sizes $\ell$ on an NVIDIA V100 GPU during inference.}
\label{time_memory}
\end{figure}

\cref{time_memory} illustrates the resource consumption under various settings, where $\ell$ denotes the size of the mini-context, and Inf represents the maximum memory usage for a given mini-context size $\ell$. When $\ell=1$, memory consumption is limited to 7.35GB, and the model operates in a purely sequential manner, substantially reducing memory demands and simplifying deployment. Increasing $\ell$ results in higher memory consumption but reduces computation time. Crucially, all settings yield consistent results, demonstrating the deployment flexibility of our model.

Supplemental \cref{tab:model_time} presents a comparison of inference times across different models. Despite employing the most extensive context (8 $\times$ 128 2D slices), our model demonstrates superior inference speed compared to other ICL models. This efficiency stems from the fact that 2D models process images slice-by-slice, necessitating the recomputation of context for each slice, thereby resulting in redundant context feature computations and prolonged inference durations. This further underscore the importance of developing Neuroverse3D which inherently avoid this issue.

\subsection{Ablation Study of Loss Functions}
We evaluated the model’s performance without the proposed modified $\mathcal{L}_{\text{smooth}-L_{1}}$ loss and gradient loss, as shown in \cref{tab:ablation}. 

Removing the modified $\mathcal{L}_{\text{smooth}-L_{1}}$ loss led to a decline in segmentation performance, particularly in small, complex regions such as the hippocampus (see supplemental \cref{tab:ablation_study} for details). This indicates that the modified $\mathcal{L}_{\text{smooth}-L_{1}}$ loss helps the model focus on more challenging regions, achieving a better balance in performance across various segmentation tasks.

Excluding the gradient loss resulted in a marked decrease in generation task performance, underscoring the importance of capturing edge information in brain images. Interestingly, even though the gradient loss was not applied to segmentation tasks, it improved segmentation performance. This improvement likely stems from the model’s heightened sensitivity to boundaries.

\begin{table}[h!]
    \centering
    \renewcommand{\arraystretch}{1}
    \setlength{\tabcolsep}{4pt}
    \small % Use small font
    \resizebox{0.47\textwidth}{!}{
    \begin{tabular}{cccc}
        \toprule
        \textbf{Modified L1} & \textbf{Gradient Loss} & \renewcommand{\arraystretch}{0.8}\begin{tabular}[x]{@{}c@{}}\textbf{Segmentation}\\\textbf{(Dice)}\end{tabular}  & \renewcommand{\arraystretch}{0.8}\begin{tabular}[x]{@{}c@{}}\textbf{Generation}\\\textbf{(PSNR)}\end{tabular}\\
        \midrule
        & \checkmark  & 0.7977 $\pm$ 0.0313 & 28.19 $\pm$ 0.65 \\
        \checkmark &  & 0.8014 $\pm$ 0.0304 & 26.67 $\pm$ 0.69 \\
        \checkmark & \checkmark & \textbf{0.8173 $\pm$ 0.0271} & \textbf{28.38 $\pm$ 0.76} \\ 
        \bottomrule
    \end{tabular}
    }
    \caption{Ablation study of loss functions demonstrating the model performance across all segmentation and generation tasks, with results averaged over context sizes of 1, 2, 4, and 8.}
    \label{tab:ablation}
\end{table}

\section{Conclusion}
This paper introduces Neuroverse3D, the first 3D In-Context Learning (ICL) universal model for neuroimaging, addressing the critical challenge of high memory consumption inherent in ICL. The proposed APSP and U-shaped fusion enable adaptive parallel-sequential context processing, supporting unlimited context. An optimized loss function balances performance across diverse tasks and enhances anatomical focus. Evaluated on cross-center data, Neuroverse3D significantly outperforms other ICL models in all tasks, matching the performance of fully supervised models in segmentation. Trained on large datasets, Neuroverse3D demonstrates robust generalization and eliminates retraining, showing strong practical potential. By overcoming memory limitations for the ICL model, Neuroverse3D fundamentally paves the way for more diverse scenarios of ICL exploration such as processing other organs in 3D.
%% Conclusion 还要再更加insightful

\section*{Acknowledgements}
This work was supported in part by grants from the National Natural Science Foundation of P.R. China (62276081), Basic
Foundation of Shenzhen Science and Technology National Key Program (CJGJZD20230724093959002), and The Major Key Project of PCL (PCL2023A09). Yanwu acknowledge the afffliation with the International Max Planck Research School for Intelligent Systems (IMPRS-IS).

{
    \small
    \bibliographystyle{ieeenat_fullname}
    \bibliography{main}
}

\clearpage
\setcounter{page}{1}
% Reset section numbering to A, B, C...
\renewcommand{\thesection}{\Alph{section}}  % Use alphabetic section numbering (A, B, C, ...)
\setcounter{section}{0}  % Reset section counter
% \onecolumn
\onecolumn
% \maketitlesupplementary
\begin{center}
    \Large 
    \textbf{Neuroverse3D: Developing 3D In-Context Learning Universal Model in Neuroimaging}\\
    \vspace{0.5em}Supplementary Material \\
    \vspace{1.0em}
\end{center}

\section{Method}
\label{sec:method_sup}

\subsection{Gradient Equivalence}
\label{sec:Gradient_Equivalence}
Here, we prove that the expected value of the gradient using Adaptive Parallel-Sequential Processing (APSP) is the same as the full-context gradient when no mini-contexts are discarded. The full-context gradient, exemplified by the $i$-th stage output $\bar{c}_{\text{dec},j}^i$, with respect to parameters $\theta$ is defined as:
\begin{equation}
    \nabla_\theta \mathcal{L}_{\text{full}}^i = \nabla_\theta \sum_{j=1}^n \frac{\text{len}(s_j)}{L}\bar{c}_{\text{dec},j}^i,
    \label{eq:full_gradient}
\end{equation}
where $L = \sum_{j=1}^n \text{len}(s_j)$ ensures equal weighting of image-label pairs across mini-contexts. For brevity, we focus on the $i$-th stage, with the understanding that the analysis applies analogously to other stages.

In APSP, mini-contexts are randomly shuffled, and we select the last mini-context to compute the gradient. Numerically, this is equivalent to randomly selecting a mini-context with index $\pi$ from $\{1, 2, ..., n\}$ with a uniform distribution, denoted as $\pi \sim \mathcal{U}\{1, ..., n\}$. The APSP gradient using only the $\pi$-th mini-context is:
\begin{equation}
    \nabla_\theta \mathcal{L}_{\text{APSP},\pi}^i = \nabla_\theta \frac{\text{len}(s_\pi)}{L}\bar{c}_{\text{dec},\pi}^i.
\end{equation}
To ensure the expected APSP gradient matches the full-context gradient, during gradient computation we scale $\bar{c}_{\text{dec},\pi}^i$ by a factor of $n$, making $\nabla_\theta\mathcal{L}_{\text{APSP-scaled},\pi}^i = \nabla_\theta  \frac{\text{len}(s_\pi)}{L}n\bar{c}_{\text{dec},\pi}^i$ the scaled gradient for the $\pi$-th mini-context. It is important to note that the scaling factor of $n$ is exclusively applied during the gradient computation phase and is omitted during the forward computation, thus having no impact on the forward computation of the model. The expected APSP gradient with scaling is:
\begin{equation}
\begin{aligned}
    \mathbb{E}_\pi \left[ \nabla_\theta \mathcal{L}_{\text{APSP-scaled},\pi}^i \right] &= \mathbb{E}_\pi \left[  \nabla_\theta \frac{\text{len}(s_\pi)}{L} n \bar{c}_{\text{dec},\pi}^i \right] \\
    &= \sum_{j=1}^{n} P(\pi = j) \left[ n \nabla_\theta \frac{\text{len}(s_j)}{L} \bar{c}_{\text{dec},j}^i \right] \\
    &= \sum_{j=1}^{n} \frac{1}{n} \left[ n \nabla_\theta \frac{\text{len}(s_j)}{L} \bar{c}_{\text{dec},j}^i \right] \\
    &= \sum_{j=1}^{n} \nabla_\theta \frac{\text{len}(s_j)}{L} \bar{c}_{\text{dec},j}^i \\
    &= \nabla_\theta \sum_{j=1}^{n} \frac{\text{len}(s_j)}{L} \bar{c}_{\text{dec},j}^i \\
    &= \nabla_\theta \mathcal{L}_{\text{full}}^i.
\end{aligned}
\label{eq:expected_gradient_scaled}
\end{equation}
Equation~\eqref{eq:expected_gradient_scaled} shows that the expected APSP gradient equals the full-context gradient $\nabla_\theta \mathcal{L}_{\text{full}}^i$. This analysis, exemplified for the $i$-th stage, extends to all network stages and parameters.

Thus, by employing the gradient of only the last mini-context and scaling by $n$, APSP preserves the full-context gradient expectation while enabling memory-efficient computation.

\subsection{U-Shape Fusion Strategy Explanation}
\label{sec:U_Shape}

\begin{figure*}[h!]
\centering
\includegraphics[width=0.7\textwidth]{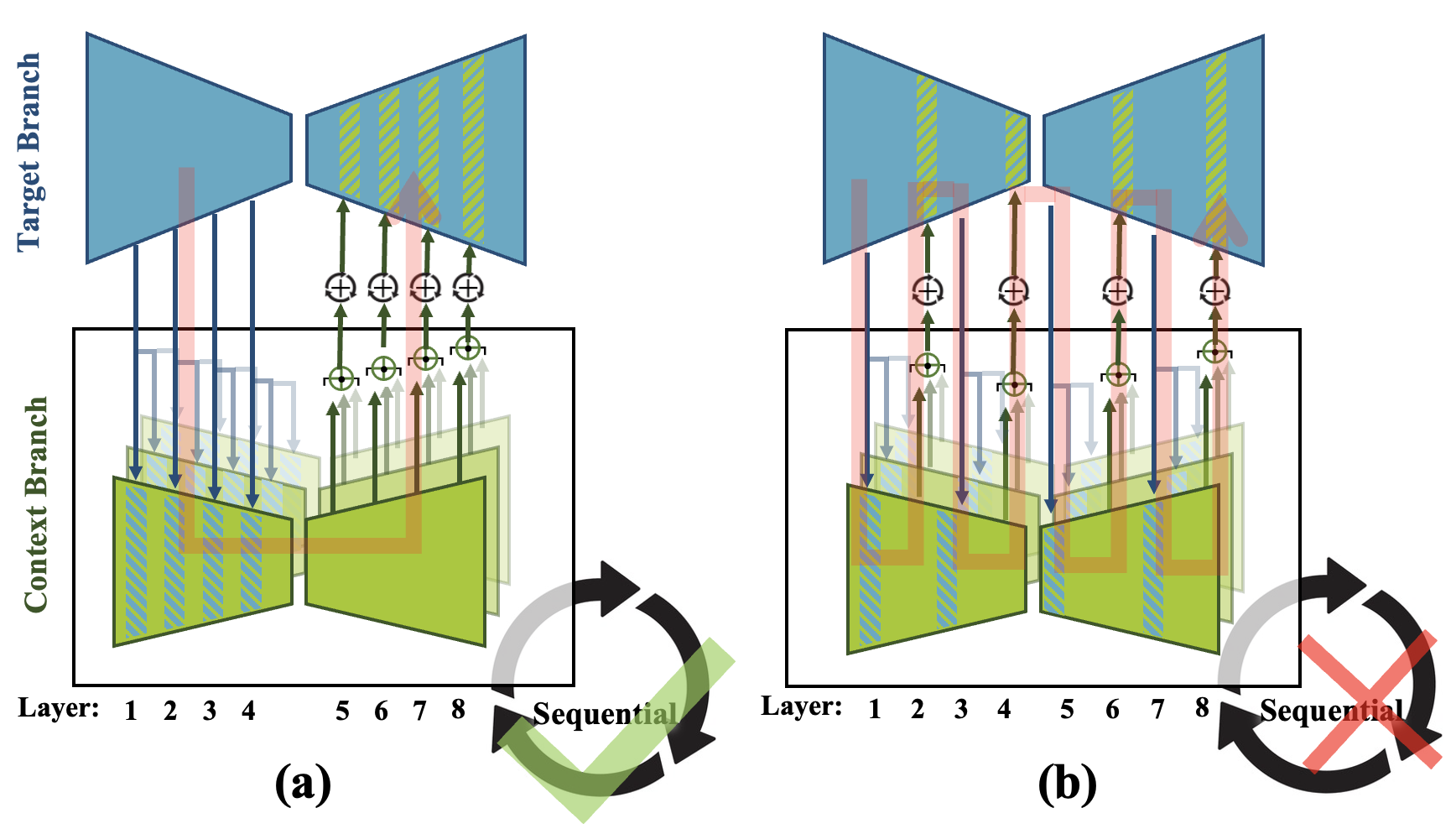}
\caption{Illustration of different fusion strategies. (a) The proposed U-Shape fusion strategy. (b) The alternating fusion strategy. The red semi-transparent arrows indicate the computation order of the feature maps in the network.}
\label{fig:ushape}
\end{figure*}

In \Cref{fig:ushape}, (a) illustrates our proposed U-Shape fusion strategy, while (b) shows a fusion strategy with alternating feature transmission, similar to those presented in \cite{butoi2023universeg} and \cite{czolbe2023neuralizer} for comparison in the following analysis. The red semi-transparent arrows indicate the order of computation for feature maps in the network.

In case (b), there is a problematic trade-off between memory usage and computational cost. For example, when computing layer 2 of the target branch, it is necessary to calculate the feature maps for all image-label pairs in the context branch at layer 2, which are then passed to the target branch. After this, two options arise:
\begin{enumerate}
    \item Retaining the context branch layer 2 features for subsequent computations requires storing feature maps for all image-label pairs across all mini-contexts during sequential processing, which significantly increases memory usage.
    \item Discarding the context branch layer 2 features saves memory but requires recomputing them for subsequent layers, which substantially increases computational cost.
\end{enumerate}
Both options severely hinder efficient sequential processing.
In contrast, strategy (a) avoids these issues. For each image-label pair, all required context representations are obtained in a single pass, allowing computed feature maps to be discarded during sequential processing, thereby ensuring computational efficiency.

\section{Data}
\label{sec:data_sup}
% 数据表格
\begin{table*}[h!]
    \centering
    \renewcommand{\arraystretch}{1.1}
    \setlength{\tabcolsep}{4pt}
    \resizebox{0.75\textwidth}{!}{
    \begin{tabular}{cccccc>{\centering\arraybackslash}p{3cm}}
        \toprule
        \textbf{Type for use} & \textbf{Dataset} & \textbf{Task} & \textbf{\# Scans} & \textbf{\# Masks} & \textbf{Modality} \\
        \midrule
        % \textbf{Training Set} & & & & & \\
        \multirow{17}{*}{\begin{tabular}[x]{@{}c@{}}Training and \\Validation Set\end{tabular}} 
        & TopCow\cite{yang2023benchmarking} & Seg., Gen. & 90 & 90 & MRA \\
        & CAS2023\cite{CAS2023} & Seg., Gen.  & 100 & 100 & MRA \\
        & ISLES2022\cite{hernandez2022isles} & Gen., Mod.  & 750 & 0 & \renewcommand{\arraystretch}{0.8}\begin{tabular}[x]{@{}c@{}}DWI, ADC,\\FLAIR\end{tabular}\\
        & ATLAS\cite{liew2022large} & Seg., Gen. & 655 & 655 & T1w \\
        & IXI\cite{IXI} & Gen., Mod.  & 2268 & 0 &   \renewcommand{\arraystretch}{0.8}\begin{tabular}[x]{@{}c@{}}T1, T2,\\MRA, PD\end{tabular}\\
        & ICH Unlabeled\cite{flanders2020construction} & Gen.  & 2000 & 0 & CT \\
        & ADHD\cite{ADHD} & Gen.  & 950 & 0 & T1 \\
        & ADNI\cite{jack2008alzheimer} & Gen., Mod. & 9923 & 0 & T1 \\
        & CMI\cite{alexander2017open} & Gen.  & 5146 & 0 & T1 \\
        & GSP\cite{holmes2015brain} & Gen.  & 2616 & 0 & T1 \\
        & HAB\cite{gera2023characterizing} & Seg., Gen.  & 460 & 460 & T1 \\
        & NIMH\cite{nugent2022nimh} & Seg., Gen.  & 248 & 248 & T1 \\
        & OASIS\cite{marcus2007open} & Gen.  & 3916 & 828 & T1 \\
        & UKBiobank\cite{sudlow2015uk} & Seg., Gen.  & 4000 & 2000 & T1, T2 \\
        & BraTS\cite{menze2014multimodal} & Seg., Gen., Mod.  & 5004 & 1251 & \renewcommand{\arraystretch}{0.8}\begin{tabular}[x]{@{}c@{}}FLAIR, T1,\\T1CE, T2\end{tabular} \\
        
        \midrule
        % \textbf{Held-out Set} & & & & & \\
        \multirow{4}{*}{Held-out Set} 
        & WMH\cite{kuijf2019standardized} & Mod. & 120 & 0 & T1, FLAIR \\
        & CCNP\cite{liu2021chinese} & Gen.  & 1580 & 0 & T1 \\
        & FCON1000\cite{FCON} & Seg., Gen.  & 1096 & 1096 & T1 \\
        & PPMI\cite{marek2011parkinson} & Gen.  & 2752 & 0 & T1 \\
        \midrule

        % & \textbf{Total} & Seg., Gen. & 38074 & 44677 & 6728 & T1, T2, FLAIR, MRA, DWI, ADC, PD, CT \\
        & \textbf{Total} & Seg., Gen., Mod.  & 43674 & 6728 & \renewcommand{\arraystretch}{0.8}\begin{tabular}[x]{@{}c@{}}T1, T2, FLAIR,\\MRA, DWI, ADC,\\ PD, CT\end{tabular} \\
        
        \bottomrule
    \end{tabular}}
    \caption{Summary of Datasets. Seg., Gen., and Mod. represent segmentation, generation (excluding modality transformation), and modality transformation tasks, respectively.}
    \label{tab:dataset_summary}
\end{table*}

\subsection{Domain Shift Between Training and Test Data}
We trained a discriminator to distinguish between images from the training set and those from the held-out set, achieving accuracy, precision, recall, and F1 scores of 0.934, 0.932, 0.918, and 0.925, respectively. Additionally, we trained a nnUNet model for brain anatomical segmentation using the training set. These results serve as proxy evidence for the presence of domain shift.

\subsection{Data Sampling}
\label{sec:data_sup_}
During training, tasks are selected based on a predefined sampling rate, followed by the random selection of a dataset within the chosen task with equal probability. Given a context size $L$, we randomly sample $L+1$ image-label pairs from the training set of the selected dataset. One of these samples is alternately designated as the target image and ground truth, while the remaining $L$ samples serve as the context set for the model. This process generates $L+1$ unique target image and corresponding context set, significantly reducing the time consumption associated with data I/O. 

Since the model is designed for binary classification, for multi-class segmentation datasets, we iteratively select each class as the foreground and the remaining classes as the background during training.

In generation tasks, we simulate various scenarios. For bias field correction, 3D bias fields are generated using Legendre polynomials with random coefficients. Gaussian noise removal involves simulating noise with a mean of 0 and a standard deviation randomly selected within the range 0.15 to 0.25. For salt-and-pepper noise removal, noise is applied with random probabilities equal to 0.04, where salt noise (value of 1) and pepper noise (value of 0) are added separately. For the inpainting task, binary masks are created using random 3D Perlin noise to occlude specific regions of the input image. The 2D-to-3D task requires the model to reconstruct a complete 3D brain volume from only three central brain slices, with this task restricted to images in MNI space. In the super-resolution task, images are downsampled by a factor of 2. For skull stripping, input images include the skull, while ground truth images, with the skull removed, are generated using FreeSurfer~\cite{fischl2012freesurfer}. Lastly, the modality transformation task uses registered pairs of different imaging modalities as input and output.

\begin{figure*}
\centering
\includegraphics[width=0.99\textwidth]{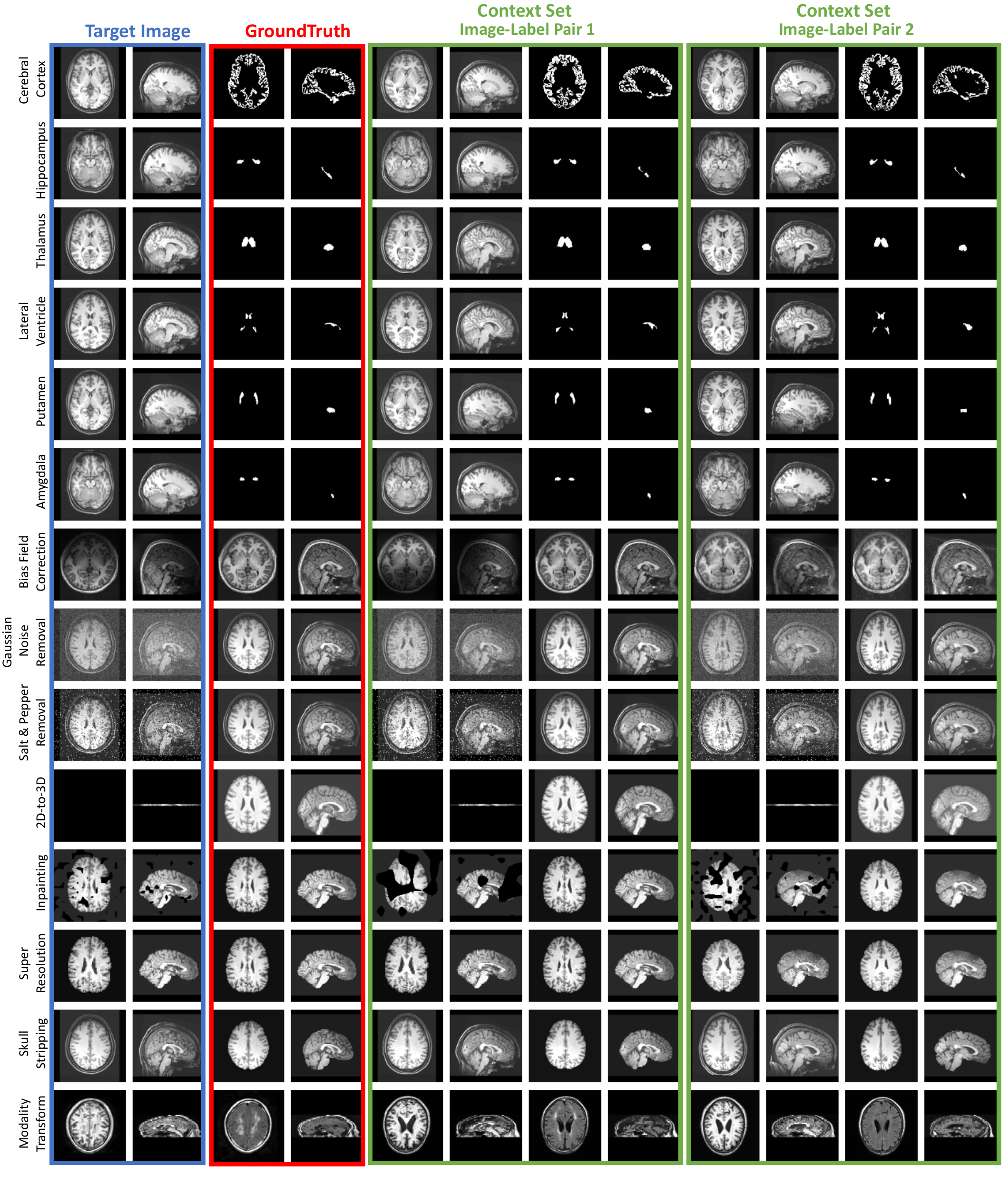}
\caption{Visualization of the target image, ground truth, and image-label pairs in the context set.}
\label{all_tasks}
\end{figure*}

\Cref{all_tasks} illustrates the target image, ground truth, and context set (with two image-label pairs shown as examples) for all tasks. The model takes the target image and context set as input, using the context set to infer the required task.

\begin{table}[h!]
\centering
\renewcommand{\arraystretch}{1} % Adjust row height
\resizebox{0.47\textwidth}{!}{
    \begin{tabular}{ccc}
    \toprule
    \textbf{Task} & \textbf{Sampled Rate} & \textbf{Weight} \\
    \midrule
    Segmentation & 2 & 50 \\
    Bias Remove & 1 & 1 \\
    Gaussian Noise Remove & 1 & 1 \\
    Salt \& Pepper Noise Remove & 1 & 1 \\
    2D to 3D Generation & 1 & 0.5\\
    Inpainting & 1 & 1 \\
    Super-Resolution & 1 & 1 \\
    Skull Stripping & 1 & 1 \\
    Modality Transform & 1 & 1 \\
    \bottomrule
    \end{tabular}
    }
    \caption{Sampling rate and weight assigned to each task when training.}
\label{tab:task_sampled_weight}
\end{table}

\begin{figure*}
\centering
\includegraphics[width=0.8\textwidth]{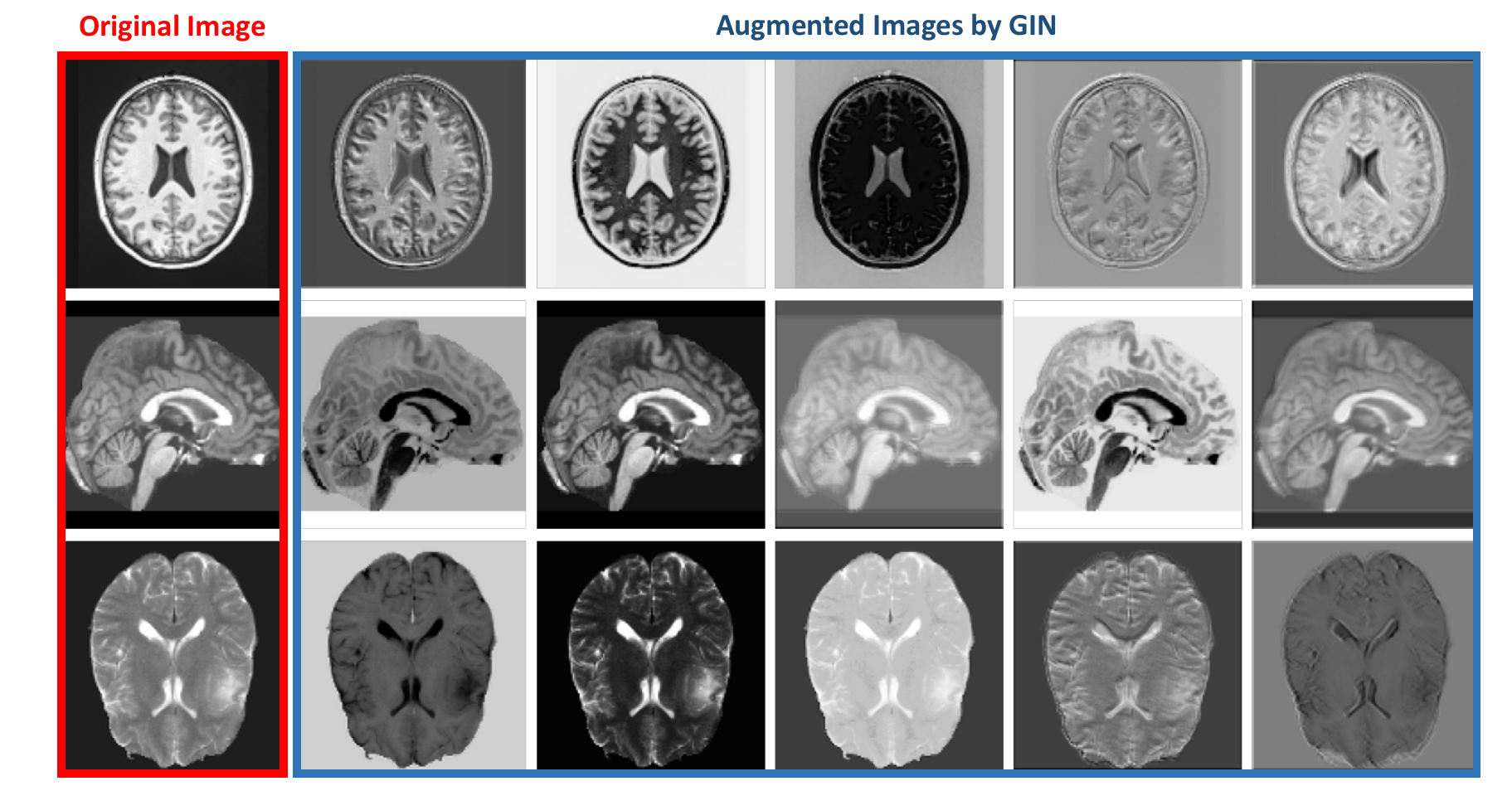}
\caption{Visualization of the images augmented by GIN~\cite{ouyang2022causality}.}
\label{gin_examples}
\end{figure*}

\begin{figure*}
\centering
\includegraphics[width=0.95\textwidth]{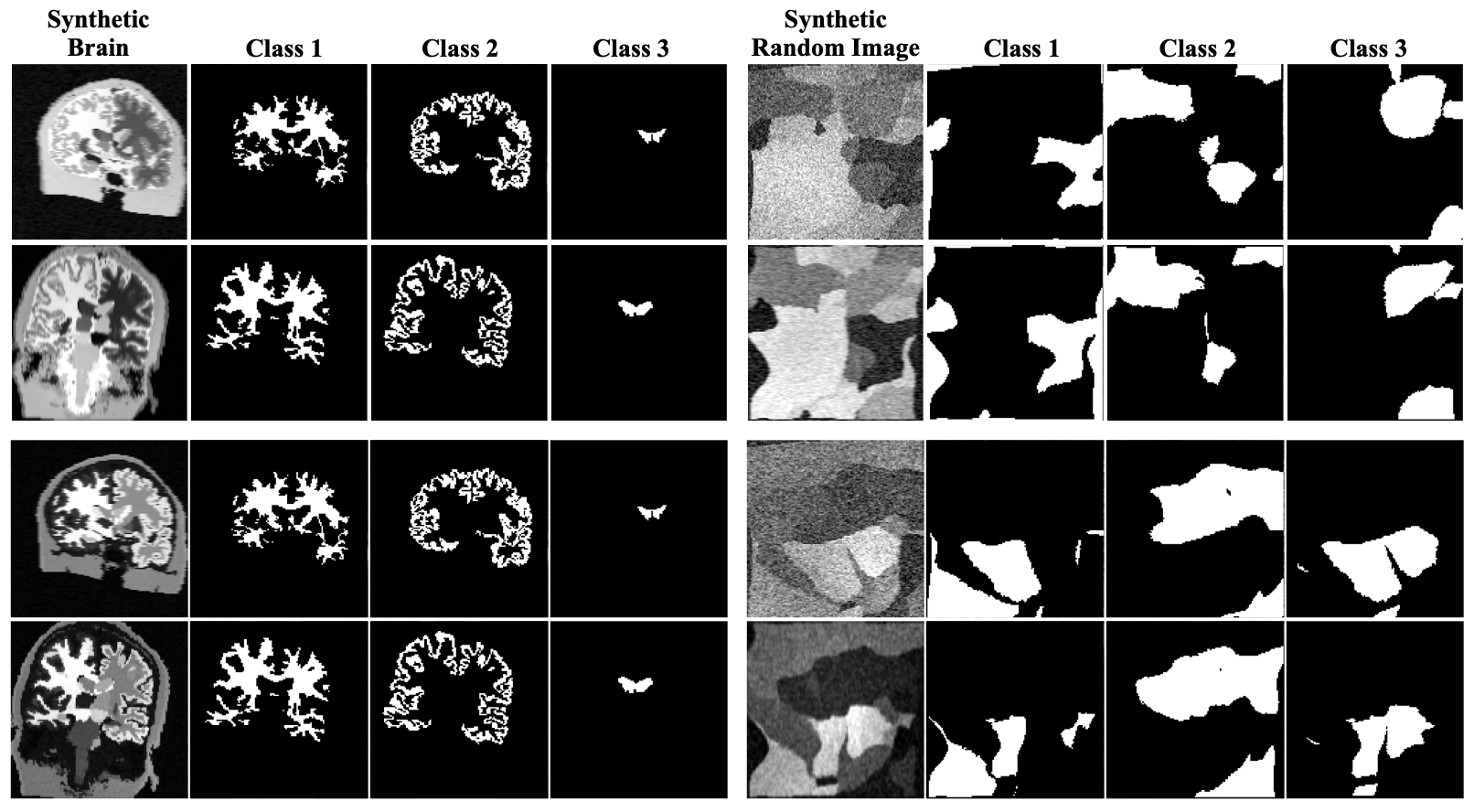}
\caption{Visualization of the synthetic images~\cite{butoi2023universeg, hoffmann2021synthmorph} for the segmentation task and corresponding segmentation masks.}
\label{syn_examples}
\end{figure*}

% \clearpage
\subsection{Image and Task Augmentation}
\label{sec:aug}

Image and task augmentation are performed after obtaining data sampling. For image augmentation, we applied the following transformations: random affine transformations (p=0.05), elastic deformations (p=0.05), flips (p=0.05), and rotations (p=0.05). To extend the diversity of input images, we introduced random intensity shifts (p=0.2), intensity scaling (p=0.2), Gaussian noise (p=0.1), intensity inversion (p=0.05), and contrast enhancement using a random convolutional network called GIN~\cite{ouyang2022causality} (p=0.05). \Cref{gin_examples} showcases examples of applying GIN to process images, resulting in the generation of random and diverse modalities.

For task augmentation, we employed several methods as follows:

\begin{itemize}
    \item \textbf{Random Task Overlapping}: To enhance the model's ability to perform multiple tasks in a single run, we randomly overlay tasks during training: bias field correction, Gaussian noise removal, salt and pepper noise removal, inpainting, and super-resolution, each with a probability of 0.05. These overlays are independent.
    \item \textbf{Random Modalities} (p=0.05): In modality transformation tasks, we also applied GIN~\cite{ouyang2022causality} to the target and ground truth images. The same parameters were used for both context and target images.
    \item \textbf{Random Foreground} (p=0.5): In multi-class segmentation datasets, multiple classes were randomly grouped into a single foreground class, with remaining classes assigned as background. This approach increases the variety of segmentation tasks. Specifically, we randomly sampled $k$ classes in the dataset (capped at 10) to form the foreground.

    \item \textbf{Sobel Filter} (p=0.05): In segmentation tasks, the Sobel filter was applied to the labels, with the model tasked to predict the filtered output.

    \item \textbf{Mask Inversion} (p=0.05): In segmentation tasks, the foreground and background masks were swapped to introduce variability.

    \item \textbf{Random Dilation} (p=0.05): The segmentation masks of the target and context are dilated by 1 voxel.

    \item \textbf{Random Erosion} (p=0.05): The segmentation masks of the target and context are eroded by 1 voxel.

\end{itemize}

Each image and task augmentation was applied independently according to its probability, and these augmentations are mutually compatible. 

\vspace{0.4em}\noindent\textbf{Synthetic Data.} To enhance generalization, we incorporated synthetic data, following~\cite{butoi2023universeg, hoffmann2021synthmorph}, as shown in~\Cref{syn_examples}. This added 100 segmentation datasets with varying contrasts, each containing 100 3D image samples.

\section{Experiments}
\label{sec:exp_sup}

\subsection{Training}
Neuroverse3D was trained on eight NVIDIA V100 GPUs, with a batch size of 1 per GPU, using the ADAM optimizer. Training ran for 120K steps with an initial learning rate of $10^{-4}$. Validation loss was evaluated every 1.2K steps, and the learning rate was halved if no improvement was observed over 20 evaluations. To improve training efficiency, the context size and mini-context size were fixed at 3 for the first 100K steps, requiring the model to process only one mini-context. Then, the context size was uniformly selected between 1 and 8 for the final 20K steps. Total training took approximately 8 days, with the model achieving the lowest validation loss selected. All task-specific models were trained on a single GPU for 36K steps, due to the much smaller training set and faster convergence compared to the multi-task ICL model.

\subsection{Evaluation}
Performance was assessed using the Dice coefficient for segmentation tasks and the Peak Signal-to-Noise Ratio (PSNR) for generation tasks on held-out datasets. Each task and setting were evaluated 10 times with randomly selected context sets to compute the average and standard deviation, ensuring robust performance measurements.

\subsection{Qualitative Results}
\label{sec:Qualitative_Results}
\begin{figure*}
\centering
\includegraphics[width=0.99\textwidth]{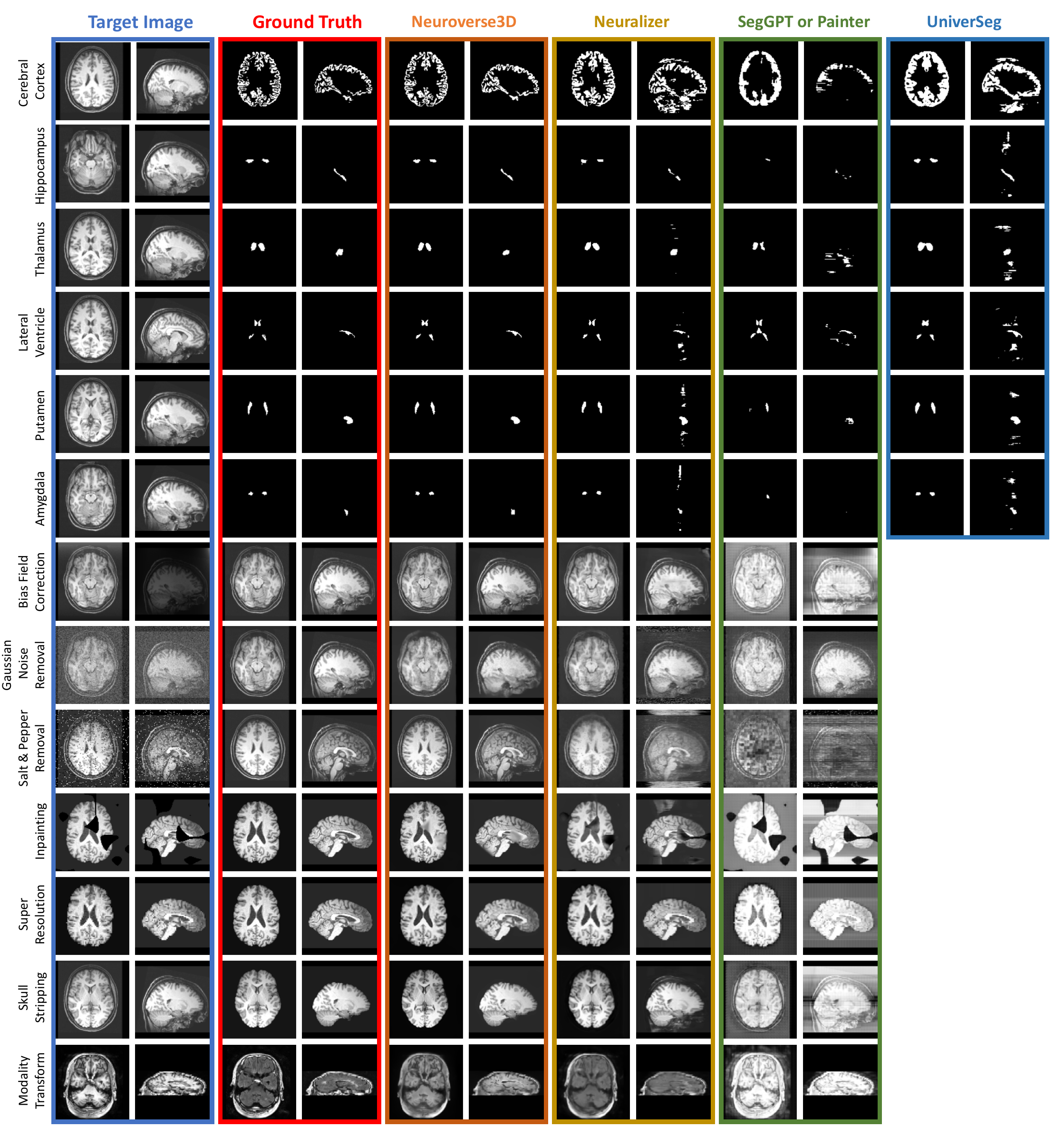}
\caption{Qualitative results comparing of ICL models. In the SegGPT or Painter column, the results for segmentation tasks are from SegGPT, and the results for generation tasks are from Painter.}
\label{comaprison1}
\end{figure*}

\Cref{comaprison1} present the comparisons between our model and other ICL models. In segmentation tasks, other ICL models often produce false positive results on background slices, while our Neuroverse3D effectively captures global information, eliminating this issue. For generation tasks, Neuroverse3D demonstrates improved slice-to-slice consistency compared to 2D ICL models, underscoring the critical importance of leveraging a 3D model.

Moreover, Painter, an ICL model trained on natural images, struggles to handle these specialized medical imaging tasks despite being exposed to extensive data and having a large number of parameters. This highlights that for ICL models, merely providing context might be insufficient for adaptation to the medical domain without incorporating domain-specific knowledge.

\begin{figure*}
\centering
\includegraphics[width=0.90\textwidth]{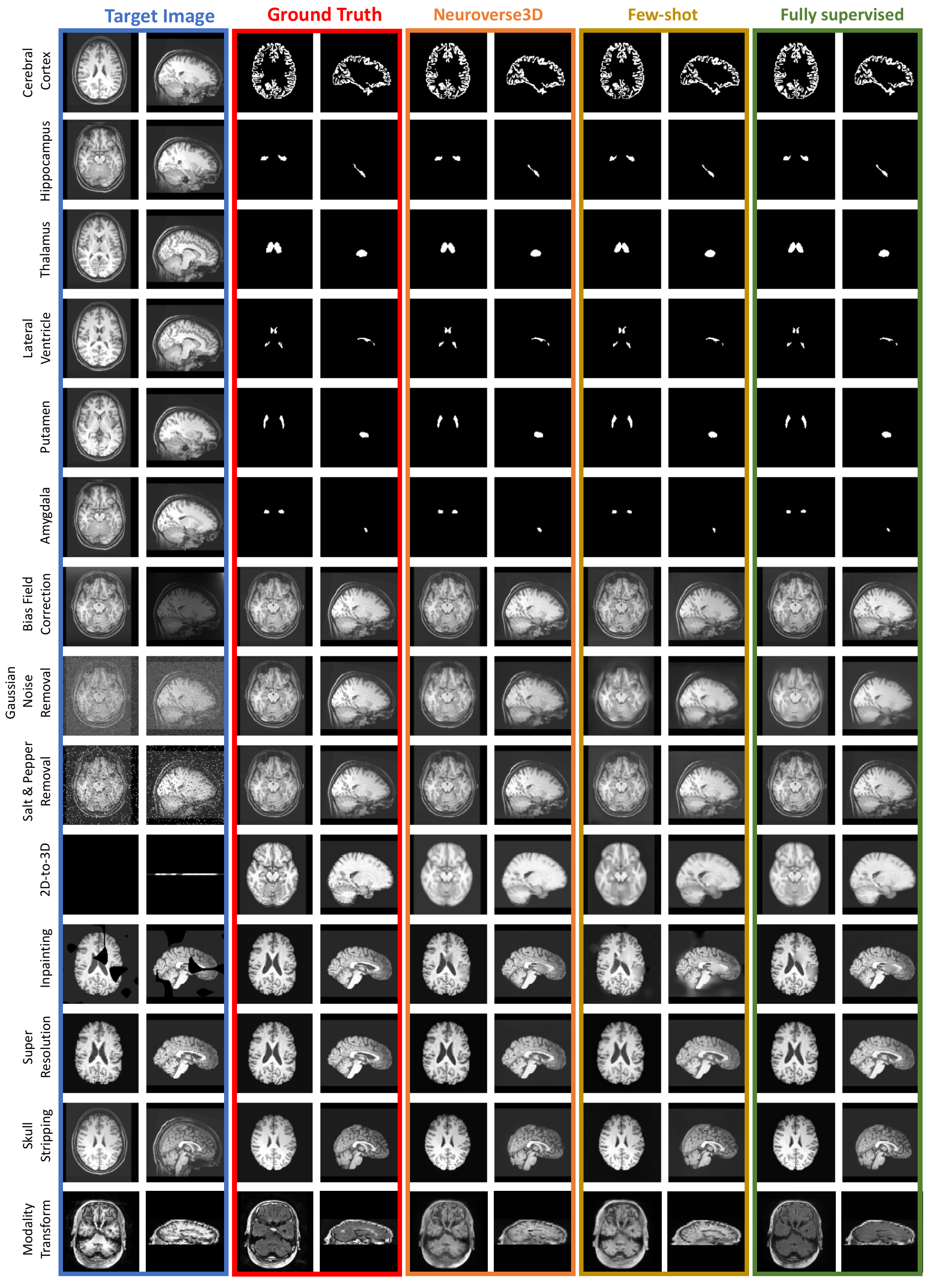}
\caption{Qualitative comparison with task-specific models.}
\label{comaprison2}
\end{figure*}

\Cref{comaprison2} presents the comparison results between our model and task-specific models. Overall, in 3D scenarios, with a limited number of samples and well designed data augmentation, few-shot task-specific models can achieve impressive results, which aligns with findings from previous studies~\cite{avesta2022comparing}. For 3D segmentation tasks, the performance of Neuroverse3D is on par with both few-shot and fully supervised models.

% 不同模型的比较

\clearpage
\begin{table*}[h]
\centering
\setlength{\tabcolsep}{2pt}
\resizebox{0.85\textwidth}{!}{
\begin{tabular}{ccccccccc}
    \toprule
\textbf{Smodified L1} & \textbf{Gradient Loss} &
\footnotesize
 \renewcommand{\arraystretch}{0.8}\begin{tabular}[x]{@{}c@{}}\textbf{Cerebral}\\\textbf{Cortex}\end{tabular}&
\footnotesize
 \textbf{Hippocampus} &
\footnotesize
 \textbf{Thalamus} &
\footnotesize
 \renewcommand{\arraystretch}{0.8}\begin{tabular}[x]{@{}c@{}}\textbf{Lateral}\\\textbf{Ventricle}\end{tabular}&
\footnotesize
 \textbf{Putamen} &
\footnotesize
 \textbf{Amygdala} &
 \textbf{Average} \\
\midrule
    & \checkmark & 0.8847	&0.7618&	0.8586&	0.8150	&0.8220&	0.6439 &0.7977\\
    \checkmark &  & 0.8724	&0.7791&	0.8648	&0.8307	&0.8167&	0.6446 &0.8014\\
    \checkmark & \checkmark & 0.8898&	0.7917&	0.8726	&0.8290	&0.8398&	0.6812& 0.8173\\
\bottomrule
\end{tabular}
}
\caption{Detailed ablation study of segmentation tasks.}
\label{tab:ablation_study}
\end{table*}

\begin{table*}[h]
\centering
\setlength{\tabcolsep}{2pt}
\resizebox{0.95\textwidth}{!}{
\begin{tabular}{ccccccccccc}
\toprule
\textbf{Smodified L1} & 
\textbf{Gradient Loss} & 
\footnotesize
\renewcommand{\arraystretch}{0.8}\begin{tabular}[x]{@{}c@{}}\textbf{Bias Field}\\\textbf{Correction}\end{tabular} & 
\footnotesize
\renewcommand{\arraystretch}{0.8}\begin{tabular}[x]{@{}c@{}}\textbf{Gaussian}\\\textbf{Noise Removal}\end{tabular} & 
\footnotesize
\renewcommand{\arraystretch}{0.8}\begin{tabular}[x]{@{}c@{}}\textbf{Salt-and-Pepper}\\\textbf{Noise Removal}\end{tabular} & 
\footnotesize
\textbf{2D-to-3D} & 
\footnotesize
\textbf{Inpainting}  & 
\footnotesize
\renewcommand{\arraystretch}{0.8}\begin{tabular}[x]{@{}c@{}}\textbf{Super}\\\textbf{Resolution}\end{tabular} & 
\footnotesize
\renewcommand{\arraystretch}{0.8}\begin{tabular}[x]{@{}c@{}}\textbf{Skull}\\\textbf{Stripping}\end{tabular} & 
\footnotesize
\renewcommand{\arraystretch}{0.8}\begin{tabular}[x]{@{}c@{}}\textbf{Modality}\\\textbf{Transformation}\end{tabular} &
\textbf{Average}\\
\midrule
    & \checkmark & 27.63&	25.57&	35.85&	25.50&	31.49&	27.71&	27.99&	23.79&	28.19\\
    \checkmark &  & 24.16&	24.93&	30.82&	25.55&	29.19&	27.48&	27.44&	23.81&	26.67\\
    \checkmark & \checkmark & 27.68&	25.89&	35.72&	26.08&	31.21&	28.08&	28.61&	23.74&	28.38 \\
\bottomrule
\end{tabular}}
\caption{Detailed ablation study of generation tasks.}
\label{tab:ablation_study_tasks}
\end{table*}

% \clearpage
\subsection{Inference Cost and Model Size}
\Cref{tab:model_cost} summarizes the FLOPs and parameter counts for the models during inference. The task-specific model is implemented as a 5-stage U-Net with channel sizes of (32, 64, 128, 256, 512). Similarly, both the target and context branches of Neuroverse3D utilize a 5-stage U-Net with the same channel configuration.

\begin{table}[!htpb]
\centering
\resizebox{0.50\textwidth}{!}{
\begin{tabular}{lcc}
\toprule
\textbf{Model} & \textbf{Parameters (M)} & \textbf{Inference TFLOPs} \\
\midrule
Task-specific & 35.02 & $1.71$ \\
Neuroverse3D ($L=1$) & 70.85 & $4.35$ \\
Neuroverse3D ($L=8$) & 70.85 & $16.8$ \\
\bottomrule
\end{tabular}}
\caption{Model parameters and inference FLOPs.}
\label{tab:model_cost}
\end{table}

\begin{table}[!htpb]
\centering
\setlength{\tabcolsep}{3pt} 
\resizebox{0.475\textwidth}{!}{
\begin{tabular}{ccccc}
\toprule
& Inference Time (s) & Context (pair) & Parameters (M) \\
\midrule
Neuroverse3D & 1.01 & 8 3D & 70.85 \\
Neuralizer~\cite{czolbe2023neuralizer} & 4.96 & 32 2D & 1.27 \\
UniverSeg~\cite{butoi2023universeg} & 8.36 & 64 2D & 1.18 \\
Painter~\cite{wang2023images} & 31.35 & 1 2D & 307.72 \\
SegGPT~\cite{wang2023seggpt} & 184.89 & 8 2D & 307.72 \\
\bottomrule
\end{tabular}}
\caption{Inference time for a single 3D image (128 2D slices) and the corresponding model settings on a V100 GPU. For other comparison methods, we adopt the optimal context settings reported in their respective papers, consistent with the settings in \cref{main_fig}.}
\label{tab:model_time}
\end{table}

\subsection{Different Fusion Stage Conﬁgurations}
\begin{table}[ht]
    \centering
    \scriptsize
    \setlength{\tabcolsep}{4pt}
    \renewcommand{\arraystretch}{1.0}
    \resizebox{0.6\textwidth}{!}{
    \begin{tabular}{lccccc}
        \hline
        \textbf{Fusion Stage(s)} & 5 & 5--4 & 5--3 & 5--2 & 5--1 (Ours) \\
        \hline
        Segmentation (Dice)       & 0.8019 & 0.8118 & 0.8274 & 0.8347 & \textbf{0.8446} \\
        PET Segmentation (Dice)   & 0.2870 & 0.3162 & 0.3714 & 0.4945 & \textbf{0.5314} \\
        Generation (PSNR)         & 25.82  & 26.36  & 27.41  & 27.94  & \textbf{28.87} \\
        \hline
    \end{tabular}
    }
    \caption{Performance under different fusion stage configurations. Stage 5 corresponds to the deepest stage.}
    \label{tab:stage_effect}
\end{table}

\cref{tab:stage_effect} summarizes the results of the ablation study on fusion stages. Denser connections improved the model's performance, especially on PET segmentation—an unseen modality.

\subsection{Impact of Different Context Modalities}

\begin{table}[ht]
    \centering
    \scriptsize
    \setlength{\tabcolsep}{2pt}
    \renewcommand{\arraystretch}{1.0}
    \resizebox{0.6\textwidth}{!}{
    \begin{tabular}{lcccccc}
        \hline
        Context Modality & Bias Removal & Gaussian & Salt \& Pepper & Inpainting & Super-res. & 2D to 3D \\
        \hline
        T1 & \textbf{28.32} & \textbf{26.31} & \textbf{35.98} & \textbf{31.52} & \textbf{28.29} & \textbf{26.63} \\
        T2 & 24.93 & 18.88 & 34.86 & 29.18 & 25.34 & 14.15 \\
        CT & 24.95 & 23.32 & 33.97 & 28.67 & 24.62 & 12.27 \\
        \hline
    \end{tabular}
    }
    \caption{Generation performance (PSNR) under different context modalities, with T1 as the target modality.}
    \label{tab:context_transforms}
\end{table}

\cref{tab:context_transforms} presents the impact of different context modalities. For all tasks, performance is optimal when the context modality matches the target modality. The performance drop in salt-and-pepper noise removal is relatively small, potentially because this task is less dependent on contextual information. In contrast, Gaussian noise removal and 2D-to-3D transformation show more pronounced performance degradation, likely due to their greater reliance on context to convey task-relevant knowledge. These findings underscore the importance of modality alignment for the effective performance of ICL models.

% \twocolumn

\end{document}